\documentclass[structabstract]{aa}
\usepackage{natbib}
\usepackage{txfonts}
\usepackage{graphicx}
\titlerunning{Limb darkening in spherical stellar atmospheres}
\authorrunning{H. R. Neilson \& J. B. Lester}

\begin{document}

\title{Limb darkening in spherical stellar atmospheres}

\author{Hilding R. Neilson\inst{1,3} \and John B. Lester \inst{2,3}}
\institute{
     Argelander Institute for Astronomy, University of Bonn, Auf dem Huegel 71, 53121 Bonn, Germany \\
     \email{hneilson@astro.uni-bonn.de}
 \and 
     Department of Chemical and Physical Sciences, 
     University of Toronto Mississauga,
     3359 Mississauga Road N., Mississauaga, Ontario, L5L 1C6, Canada\\
     \email{lester@astro.utoronto.ca}
 \and
     Department of Astronomy \& Astrophysics, University of Toronto,
     50 St. George Street, Toronto, Ontario, M5S 3H4, Canada
}

\date{}

\abstract
 {Stellar limb darkening, $I(\mu = \cos \theta)$, is an important constraint for microlensing, eclipsing binary, planetary transit, and interferometric observations, but is generally treated as a parameterized curve, such as a 
  linear-plus-square-root law.  Many analyses assume limb-darkening coefficients computed from model stellar atmospheres. However, previous studies, using $I(\mu)$ 
  from plane-parallel models, have found that fits to the 
  flux-normalized curves pass through a fixed point, a common $\mu$ 
  location on the stellar disk, for all values of $T_\mathrm{eff}$, 
  $\log g$ and wavelength.}
 {We study this fixed $\mu$-point to determine if it is a property 
  of the model stellar atmospheres or a property of the limb-darkening 
  laws. Furthermore, we use this limb-darkening law as a tool to probe properties of stellar atmospheres for comparison to limb-darkening observations. }
 {Intensities computed with plane-parallel and spherically-symmetric \textsc{Atlas}
  models (characterized by the three fundamental parameters $L_\star$, 
  $M_\star$ and $R_\star$) are used to reexamine the existence of the 
  fixed $\mu$-point for the parametrized curves.}
 {We find that the intensities from our spherical models do not have 
  a fixed point, although the curves do have a minimum spread at a 
  $\mu$-value similar to the parametrized curves.  We also find that 
  the parametrized curves have two fixed points, $\mu_1$ and $\mu_2$, 
  although $\mu_2$ is so close to the edge of the disk that it is 
  missed using plane-parallel atmospheres. We also find that the spherically-symmetric models appear to agree better with published microlensing observations relative to plane-parallel models.  }
 {The intensity fixed point results from the choice of the 
  parametrization used to represent the limb darkening and from the 
  correlation of the coefficients of the parametrization, which is a 
  consequence of their dependence on the angular moments of the 
  intensity.  For spherical atmospheres, the coefficients depend on the 
  three fundamental parameters of the atmospheres, meaning that limb-darkening laws contain information about stellar atmospheres. This suggests that limb-darkening parameterizations fit with spherically-symmetric model atmospheres are powerful tools for comparing to observations of red giant stars.}


\maketitle

\section{Introduction} \label{sec:intro}

The variation of the specific intensity over a star's disk, commonly 
called limb darkening or center-to-limb variation, is an important 
function of the physical structure of a stellar atmosphere.  Because it 
is difficult to observe limb darkening for stars other than the 
Sun, it is common to represent limb darkening by analytic expressions 
whose coefficients are determined by matching the $I_\lambda(\mu)$ 
from model stellar atmospheres.  Here $\mu = \cos \theta$, where 
$\theta$ is the angle between the vertical direction at that point on 
the stellar disk and the direction toward the distant observer.  
The most basic form of the law is the linear version, depending on 
$\mu$ to the first power, such as 
\begin{equation}
\frac{I_\lambda(\mu)}{I_\lambda(\mu = 1)} = \mu,
\end{equation}
or, more generally,
\begin{equation}
\frac{I_\lambda(\mu)}{I_\lambda(\mu = 1)} = 1 - u (1 - \mu).
\end{equation}
Limb-darkening laws have subsequently become more complex by including 
terms with $\mu$, or its alternative $r = \sin \theta$, raised to 
higher integer powers as well as both $\sqrt{\mu}$ and $\log(\mu)$ 
terms \citep{2000A&A...363.1081C, Howarth2010}.  In addition to minimizing the fit 
to the model's $I_\lambda(\mu)$, the laws have also introduced 
various constraints, such as enforcing flux conservation at each 
observed wavelength.  The intricacies of limb-darkening laws have 
increased as model stellar atmospheres have advanced.

\citet{2000PhDT........11H,2003ApJ...594..464H,2007ApJ...656..483H}, 
motivated by the potential of gravitational microlensing to provide 
particularly detailed measurements of stellar limb darkening, 
investigated the optimum method of extracting limb darkening from the 
data.  One curious feature that emerged from these studies is the 
existence of a fixed $\mu$ location on the stellar disk where the 
fits to the normalized model intensities have nearly the same value 
independent of wavelength or the model's $T_\mathrm{eff}$ or 
$\log g$.  This is true for both the principal-component 
analysis developed by \citet{2003ApJ...594..464H} and the more 
traditional linear limb-darkening law used by others.

\citet{2003ApJ...594..464H} used models computed with the 
\textsc{Atlas12} code \citep{1996IAUS..176..523K} to construct his 
fitting procedure.  These models use detailed opacity sampling to 
include many tens of millions of spectral lines in the radiative 
transfer, but they still assumed plane-parallel geometry, even though 
\citet{2003ApJ...594..464H} targeted red giants with 
$T_\mathrm{eff} \leq 4000$ K and $\log g \leq 1.0$.  These stellar 
parameters are exactly those where the assumption of plane-parallel 
geometry should break down.  In the study by 
\citet{2003A&A...412..241C}, which did use the spherical 
models of \citet{1999ApJ...525..871H}, the focus was on models with 
$\log g \geq 3.5$, for which spherical extension is minimal.  A 
broader study of limb darkening and the fixed $\mu$-point using 
spherically extended model atmospheres is clearly needed.

\section{\textsc{SAtlas} model atmospheres} \label{sec:satlas_models}

\citet{2008A&A...491..633L} have developed the \textsc{SAtlas} code, 
a spherically extended version of \textsc{Atlas}.  This code shares 
with \textsc{Atlas} the properties of static pressure structure, LTE 
populations and massive line blanketing represented by either opacity 
distribution functions or opacity sampling (the faster opacity 
distribution function version is used here).  The spherical program 
differs from the plane-parallel \textsc{Atlas} by allowing the gravity 
to vary with radial position, and by computing the radiative transfer 
along rays through the atmosphere in the direction of the distant 
observer using the \citet{1971JQSRT..11..589R} version of the 
\citet{1964CR...258..3189F} method, which accounts for the radial 
variation of the angle between the vertical and the direction of the 
ray.  The structure of the atmosphere is computed using a total of 81 
rays whose angular spacing is determined by two factors.  The first is 
the distribution of rays chosen to represent the ``core'' of the 
atmosphere, the region where the lower boundary condition for radiative 
transfer is the diffusion approximation.  \citet{2008A&A...491..633L} 
found that different distributions of the core rays had no effect on 
the structure of the atmosphere, and so elected to use equal spacings 
in $\mu$.  The remainder of the rays are tangent to the atmospheric 
levels at the stellar radius perpendicular to the central ray toward 
the observer.  These rays are projections of the radial spacing of the 
atmospheric levels, which is logarithmic in the Rosseland mean optical 
depth.  Because this distribution of rays is set by calculating the 
structure of the atmosphere, it is not necessarily optimal for studying 
limb darkening.  Therefore, after computing the structure of the 
atmosphere, the surface intensities are derived from the structure rays 
by cubic spline interpolation for any desired number of rays with any 
desired distribution over the disk.  As \citet{2007ApJ...656..483H} 
has demonstrated, cubic spline provides an excellent interpolation 
method.  We have created our surface intensities at 1000 points equally 
spaced over $0 \leq \mu \leq 1$.

\section{Fixed point of stellar limb-darkening laws} 
\label{sec:fixed_pt}

\citet{2000PhDT........11H} found the fixed point in the limb-darkening 
profiles of both the Sun and in plane-parallel \textsc{Atlas} models of 
red giants.  \citet{2003ApJ...596.1305F} found a similar fixed point 
in their analysis of a K3 giant using spherical \textsc{Phoenix} models,
although they excluded the limb from their intensity analysis because 
the low intensity near the limb contributed almost nothing to the 
observations they were analyzing.  Their truncation point ranged from 
$r = 0.998$ for $\log g = 3.5$, corresponding to $\mu = 0.063$, to 
$r = 0.88$ for $\log g = 0.0$, corresponding to $\mu = 0.475$.  However,
the truncation eliminated that part of the surface brightness that 
deviates most strongly from the plane-parallel model.  
\citet{2003ApJ...596.1305F} stated that the fixed point is a generic 
feature of any single-parameter limb-darkening law that conserves flux. 
For example, if the limb darkening is represented as 
$f(\mu) \equiv I(\mu)/F = 2[1 + Ax(\mu)]$, the fixed point is 
$\mu_\mathrm{fixed} = f^{-1} 
[2 \int_0^1 f(\mu') \mu' \mathrm{d} \mu']$.  
Although they also state that the fixed point is not required by the 
two-parameter law they employed, they found a fixed point in the 
microlensing observations they were modeling using such a law.

To explore the parametrization more closely, we begin with the same 
two-parameter normalized limb-darkening function used by 
\citet{2003ApJ...596.1305F}, 
\begin{equation} \label{eq:fields}
\frac{I(\mu)}{2\mathcal{H}} 
= 1 - A \left (1 - \frac{3}{2}\mu \right )
    - B \left (1 - \frac{5}{4}\sqrt{\mu} \right ),
\end{equation}
where $\mathcal{H}$ in the Eddington flux, defined as
\begin{equation}
\mathcal{H} \equiv \frac{1}{2} \int_{-1}^{1} I(\mu) \mu \mathrm{d} \mu
            = \int_{0}^{1} I(\mu) \mu \mathrm{d} \mu.
\end{equation}
A fixed point requires, for any two arbitrary models, that the 
intensity profiles satisfy
\begin{equation}
I_1(\mu_0) = I_2(\mu_0),
\end{equation}
or in terms of Eq.~\ref{eq:fields}
\begin{eqnarray} \label{eq:fields_fix}
\nonumber &&1 - A_1 \left (1 - \frac{3}{2} \mu_0 \right )
           - B_1 \left (1 - \frac{5}{4} \sqrt{\mu_0} \right ) = \\ 
&& \mbox{\hspace{1.5cm}}1 - A_2 \left (1 - \frac{3}{2} \mu_0 \right )
           - B_2 \left (1 - \frac{5}{4} \sqrt{\mu_0} \right ).
\end{eqnarray}
For this to be true, $A$ and $B$ must be linearly dependent, 
$A = \alpha B + \beta$ or $A_1-A_2 = \alpha (B_1-B_2)$.  Substituting 
this relation into Eq.~\ref{eq:fields_fix} yields an equation 
for the fixed-point 
\begin{equation} \label{eq:fp}
\frac{3}{2} \alpha \mu_0 + \frac{5}{4} \sqrt{\mu_0}  -1 - \alpha = 0.
\end{equation}
We need to verify that $A = f(B)$ and to understand the properties of 
the parameter $\alpha$.

Applying a general least-squares method to Eq.~\ref{eq:fields},
\begin{equation}
\chi^2 = \sum_i^N \left[\frac{I(\mu_i)}{2\mathcal{H}} - 1 
         + A\left(1-\frac{3}{2}\mu_i\right) 
         + B \left(1 - \frac{5}{4} \sqrt{\mu_i}\right) \right]^2,
\end{equation}
we determine the coefficients $A$ and $B$ from the constraint equations 
\begin{eqnarray} \label{eq:lsa}
 \nonumber \frac{\partial \chi^2}{\partial A}& = &
 \sum_i^N\left[\frac{I(\mu_i)}{2\mathcal{H}} - 1 
+ A\left(1-\frac{3}{2}\mu_i\right) 
+ B \left(1 - \frac{5}{4} \sqrt{\mu_i}\right) \right]  \\
&& \times    \left(1 - \frac{3}{2}\mu_i\right) = 0
\end{eqnarray}
and 
\begin{eqnarray} \label{eq:lsb}
 \nonumber \frac{\partial \chi^2}{\partial B}& =&
\sum_i^N \left[\frac{I(\mu_i)}{2\mathcal{H}} - 1 
+ A\left(1-\frac{3}{2}\mu_i\right) 
+ B \left(1 - \frac{5}{4} \sqrt{\mu_i}\right) \right] \\
&& \times    \left(1 - \frac{5}{4} \sqrt{\mu_i}\right) = 0.
\end{eqnarray}
Multiplying Eq.~\ref{eq:lsa} and Eq.~\ref{eq:lsb} by 
$\Delta \mu$ and then converting the summation to integration, 
$\int_{0}^{1} \mathrm{d} \mu$, we create the equations that determine 
$A$ and $B$,
\begin{equation}
\label{eq:lsa2}
\frac{J}{2\mathcal{H}} + \frac{1}{4}A + \frac{1}{6}B - 1 = 0
\end{equation}
and
\begin{equation} \label{eq:lsb2}
- \frac{5}{4} 
\left[\frac{1}{2\mathcal{H}} \int_0^1 I(\mu) \sqrt{\mu} \mathrm{d} \mu 
\right] + 
\frac{J}{2\mathcal{H}} + \frac{1}{6}A + \frac{11}{96}B - \frac{1}{6} 
= 0,
\end{equation}
where $J$ is the usual mean intensity.  It is also useful to define an 
angular pseudo-moment of the intensity, 
\begin{equation} \label{eq:def_P}
\mathcal{P} \equiv \int_0^1 I(\mu) \sqrt{\mu} \mathrm{d}\mu,
\end{equation}
which allows Eq.~\ref{eq:lsb2} to be written as
\begin{equation} \label{eq:lsb3}
- \frac{5}{4} \left[ \frac{\mathcal{P}}{2\mathcal{H}} \right ] + 
\frac{J}{2\mathcal{H}} + \frac{1}{6}A + \frac{11}{96}B - \frac{1}{6} 
= 0.
\end{equation}
Both $J$ and $\mathcal{P}$ are 
determined by the properties of the model stellar atmosphere or the 
star whose observations are being fit by a limb-darkening law. 
Equation~\ref{eq:lsa2} and Eq.~\ref{eq:lsb2} clearly show that the 
coefficients $A$ and $B$ are uniquely determined by $J$ and 
$\mathcal{P}$.  However, for $A$ and $B$ to be linearly dependent, 
that is $A = \alpha B + \beta$, $J$ and $\mathcal{P}$ must also be 
linearly related.

The relation of $J$ to $\mathcal{P}$ can be understood by following 
the discussion of the diffusion approximation in 
\citet{1978stat.book.....M}.  This begins by representing the source 
function by a power series, which leads to the intensity being given by 
$I_\nu(\tau_\nu, \mu) = \sum_{n = 0}^{\infty} \mu^n 
\left [\mathrm{d}^n B_\nu(T) / \mathrm{d}\tau_\nu^n \right]$
(Mihalas 1978, Eq. 2-88).  Using this expansion in the definitions of 
$J$ and $\mathcal{P}$, and keeping just the first-order term, gives the 
familiar results that $J_{\nu} \approx B_{\nu}$ and 
$K_{\nu} \approx B_{\nu}/3$ plus the additional result that 
$\mathcal{P}_{\nu} \approx 2 B_{\nu}/3$.  Eliminating $B_{\nu}$ between 
the $J_{\nu}$ and $\mathcal{P}_{\nu}$ expressions gives 
\begin{equation} \label{eq:P_J}
\mathcal{P}_{\nu} = 2J_{\nu}/3,
\end{equation}
confirming the desired correlation.  Of course, as the atmosphere thins 
out toward space the diffusion approximation becomes less accurate, but, as will be seen, 
the basic correlation of $\mathcal{P}$ and $J$ is still there.

Using Eq.~\ref{eq:P_J} we can combine Eq.~\ref{eq:lsa2} and 
Eq.~\ref{eq:lsb2} to find that
\begin{equation} \label{eq:a_fn_b}
A = -0.694 B.
\end{equation}
In terms of the notation used earlier, $\alpha = -0.694$ and 
$\beta = 0$, or equivalently $\Delta A = -0.694 \Delta B$.  Using this 
value for $\alpha$ in Eq.~\ref{eq:fp} leads to the equation for the 
fixed point, 
\begin{equation}
1.084 \mu_0^2 - 0.925 \mu_0 + 0.094 = 0.
\end{equation}
This quadratic equation yields \emph{two} fixed points, not one.  The 
solutions are $\mu_1 = 0.736$, corresponding to 
$\theta_1 = 42.61^{\degr}$ and $r_1 = 0.677$, and $\mu_2 = 0.118$, 
corresponding to $\theta_2 = 83.22^{\degr}$ and $r_2 = 0.993$.

\begin{figure} 
  \resizebox{\hsize}{!}{\includegraphics{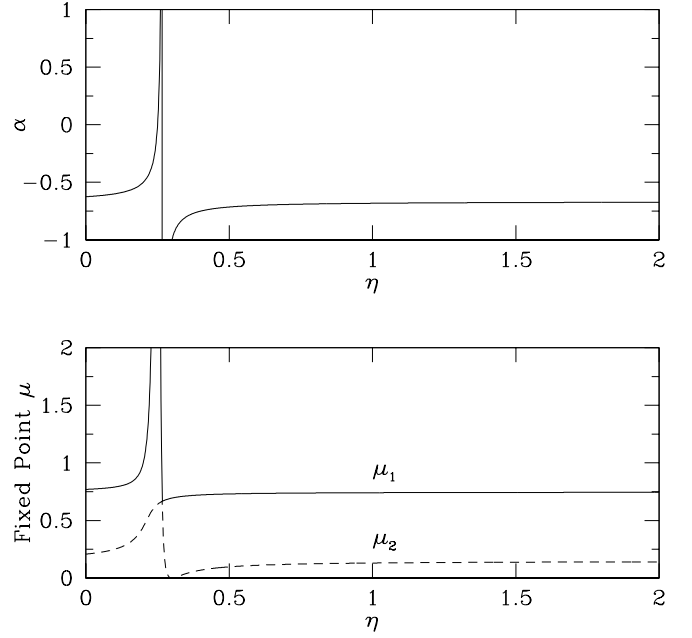}}
  \caption{Top panel: dependence of the slope of the function 
           $A = \alpha B + \beta$ on the variable 
           $\eta = J^{-1} \int I \sqrt{\mu} \mathrm{d} \mu$.
           Bottom panel: variation of the fixed points with
           $\eta$.
           \label{fig:alpha_mu_eta}
          }
\end{figure}

The results above used the diffusion approximation to establish the 
relationship between $J$ and $\mathcal{P}$.  To generalize these 
results, we define a new variable, $\eta$, as 
\begin{equation} \label{eq:eta_def}
\eta \equiv \frac{\int I \sqrt{\mu} \mathrm{d} \mu}{J} 
          = \frac{\mathcal{P}}{J}.
\end{equation}
Note that $\eta$ is a generalization of the relation in 
Eq.~\ref{eq:P_J}.  Using the definition of $\eta$, we replace 
the variable $\mathcal{P}$ in Eq.~\ref{eq:lsb3} by $\eta J$, and 
then Eq.~\ref{eq:lsa2} is used to eliminate the $J/2\mathcal{H}$ 
terms.  The resulting equation is rearranged to the form $A = f(B)$,
which is a linear equation of the same form as the previous equation 
for $A$, namely $A = \alpha B + \beta$.  Comparing the two equations 
for $A$ we identify 
\begin{equation} \label{eq:alpha}
\alpha = - \frac{(5 \eta / 4 - 1)/6 + 11/96}
              {(5 \eta / 4 - 1)/4 + 1/6}.
\end{equation}
Equation~\ref{eq:alpha} enables us to explore how $\alpha$ 
changes with $\eta$.  In addition, because Eq~\ref{eq:fp} is a 
function of $\alpha$ only, we are also able to determine the 
dependence of the two fixed points, $\mu_1$ and $\mu_2$, on $\eta$.
These dependencies are shown in Fig.~\ref{fig:alpha_mu_eta}.  Because 
the value of $\eta$ can be set to values other than 2/3, the 
assumption of the diffusion approximation is no longer being used.

Figure~\ref{fig:alpha_mu_eta} shows that, except for 
$\eta \approx 0.267$, the value of $\alpha$ is in the narrow range 
$-0.72 \lesssim \alpha \lesssim -0.68$, which leads to the fixed points 
being $0.72 \lesssim \mu_1 \lesssim 0.74$ and 
$0.08 \lesssim \mu_2 \lesssim 0.14$.  The divergence of $\alpha$ to 
$\pm \infty$ as $\eta$ approaches $\approx 0.267$ is due to the 
denominator of Eq.~\ref{eq:alpha} approaching zero as 
$\eta \rightarrow 4/15$.  However, $\eta$ can never equal $4/15$.  This 
is a consequence of the following inequalities, which are true because 
$\mu \leq 1$:
\begin{equation}
J = \int I \mathrm{d} \mu \geq \int I \sqrt{\mu} \mathrm{d} \mu 
                           \geq \int I \mu^2 \mathrm{d} \mu = K.
\end{equation}
Dividing through by $J$ this becomes
\begin{equation} \label{eq:eta_lims}
1 \geq \eta \geq \frac{K}{J}.
\end{equation}
Deep in the atmosphere, where the diffusion approximation holds, 
$K/J = 1/3$.  Moving toward the surface the atmosphere becomes more 
transparent and the radiation becomes less isotropic, with the 
consequence that $K/J > 1/3$.  As a result, 
\begin{equation} \label{eq:eta_llim}
\eta \geq 1/3 > 4/15 = 0.267,
\end{equation}
and the divergence cannot occur.  Equation~\ref{eq:alpha} can be used 
to set bounds on $\alpha$ by using the lower bound on $\eta$ from 
Eq.~\ref{eq:eta_llim} and the upper bound on $\eta$ from 
Eq.~\ref{eq:eta_lims}.  The result is  $-5/6 \leq \alpha \leq -15/22$.
Using this range of $\alpha$ in Eq.~\ref{eq:fp}, we find that the 
ranges for the fixed points are $0.708 \leq \mu_1 \leq 0.741$ and 
$0.025 \leq \mu_2 \leq 0.130$, which are consistent the values for 
$\mu_1$ and $\mu_2$ found previously.  This indicates that the fixed 
points exist in the limb-darkening relations and are stable to 
differences in the intensity profile of the stellar atmosphere.  It 
also suggests that the fixed points are not a result of a particular 
dominant opacity source.  The only requirement is that $\eta > 1/3$.

\begin{figure}[t]
  \begin{center}
     \resizebox{\hsize}{!}{\includegraphics{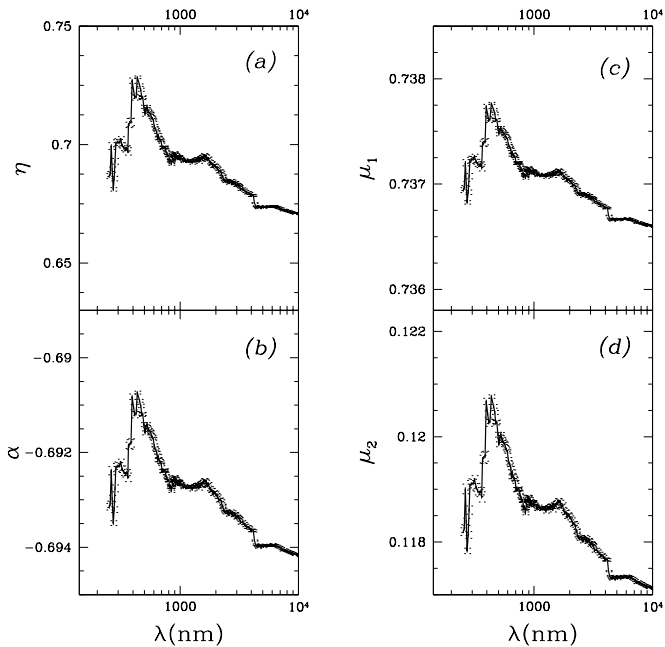}}
  \end{center}
  \caption{Wavelength dependence of the mean values of 
           (a) the ratio of the pseudo-moment to the mean intensity,
               $\eta$,
           (b) the value of $\alpha$ that defines the fixed points,
           (c) the first and
           (d) second fixed point for the grid of plane-parallel 
               \textsc{Atlas} model atmospheres. The error bars 
               represent 1-$\sigma$ deviation from the mean.
           \label{fig:wl_pp}
          }
\end{figure}

We verify this analytic result by computing a grid of \textsc{Atlas}
plane-parallel model stellar atmospheres with 
$3000 \ \mathrm{K} \leq T_\mathrm{eff} \leq 8000 \ \mathrm{K}$ and  
$-2 \leq \log g \leq 3$.  For this grid of models, the mean value of 
$\eta$ in the $V-$band is 
$\bar{\eta}_\mathrm{pp} = 0.716\pm 0.004$ which leads to 
$\alpha_\mathrm{pp} = 0.691 \pm 0.001$, $\mu_1 = 0.7375\pm 0.0002$ and
$\mu_2 = 0.1200\pm 0.0003$.  This is very similar to what is found 
assuming the diffusion approximation.  The difference between the model 
$V-$band $\eta$ and the value of $\eta$ from the diffusion 
approximation is due to the fact that plane-parallel models do not 
enforce the diffusion approximation at all depths in the atmosphere.  
To determine the dependence on wavelength, we used the plane-parallel 
grid of atmospheres to determine the mean value and 1-$\sigma$ standard 
deviation of $\eta$, $\alpha$ and the two fixed points $\mu_{1}$ and 
$\mu_{2}$, with the results shown in Fig.~\ref{fig:wl_pp}. The values 
of the fixed points clearly do not vary significantly as a function of 
wavelength nor as a function of effective temperature and gravity.  
Furthermore, it must be noted that the value of $\eta$ approaches $2/3$ 
as $\lambda \rightarrow \infty$.

These results can be generalized to show that the fixed point occurs in 
other limb-darkening laws, such as a quadratic law where
$I/2\mathcal{H} = 1 - A(1 - \mu) - B(1 -2 \mu^2)$. Repeating the 
derivation above, the pseudo-moment, $\mathcal{P}$, is replaced by $K$, 
a higher angular moment of the intensity.  Because the diffusion 
approximation analysis also yields a linear relation between $J$ and 
$K$, $J = 3K$, we again find that $A = f(B)$.  For a more general law, 
such as $I/2\mathcal{H} = 1 - A (1 - \mu) - B[1 - \frac{1}{2}(n+2)\mu^n]$, we would 
find a similar connection between the coefficients $A$ and $B$.  We 
would even find fixed points for any law that is a combination of a 
linear term and any function that can be represented by a power-law 
series, such as $e^{\mu}$, similar to the limb-darkening law tested by \cite{2003A&A...412..241C}.  Therefore, we conclude that the fixed 
points are a general property of this family of limb-darkening laws.

\section{Limb darkening and \textsc{SAtlas} model atmospheres}
\label{sec:ld_satlas}

\begin{figure}
   \resizebox{\hsize}{!}{\includegraphics{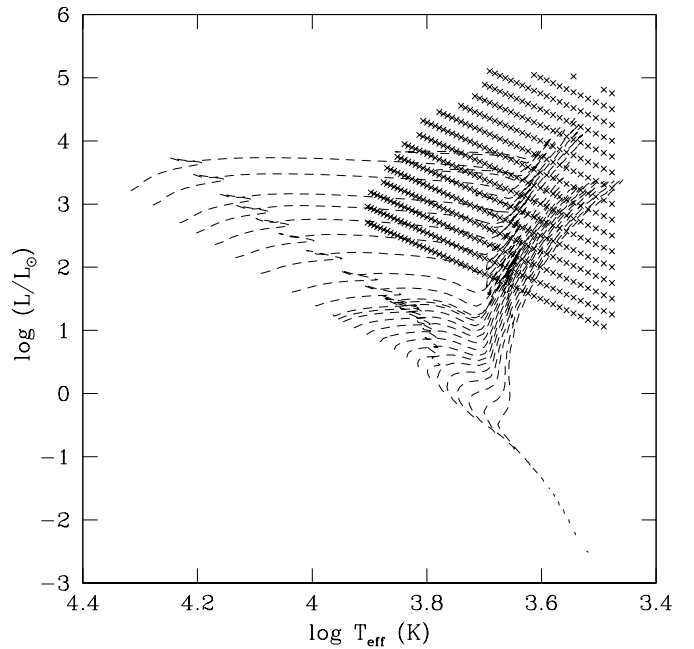}}
   \caption{Models computed with the \textsc{SAtlas} code plotted over 
            the stellar evolution tracks from 
            \citet{2000A&AS..141..371G}.
            \label{fig:ld_hr}
           }
\end{figure}

To conduct a broader survey of limb darkening, we have computed several 
larger cubes of solar-composition model atmospheres, increasing the 
range of luminosity, mass and radius as well as using microturbulent 
velocities of 0, 2 and 4 km/s.  The luminosities, masses and radii of 
these models cover a significant portion of the Hertzsprung-Russell 
diagram where stars have extended atmospheres, as shown in 
Fig.~\ref{fig:ld_hr}, which also shows the evolutionary tracks of 
\citet{2000A&AS..141..371G} for comparison.
\begin{figure*}[t]
   \resizebox{\hsize}{!}{\includegraphics{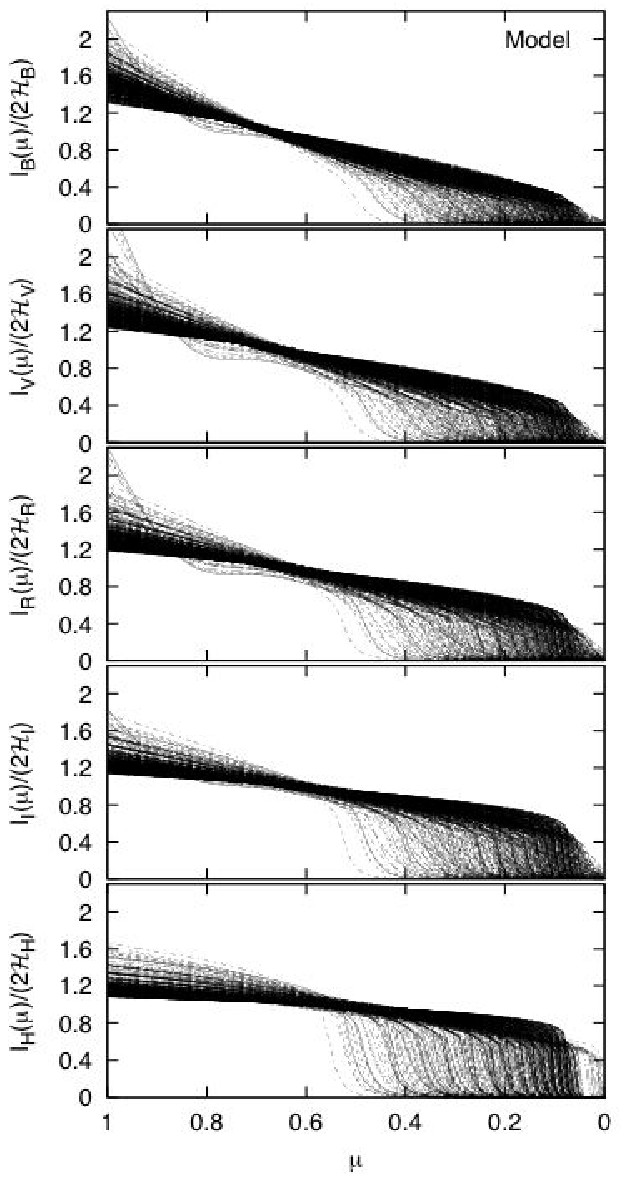}
                         \includegraphics{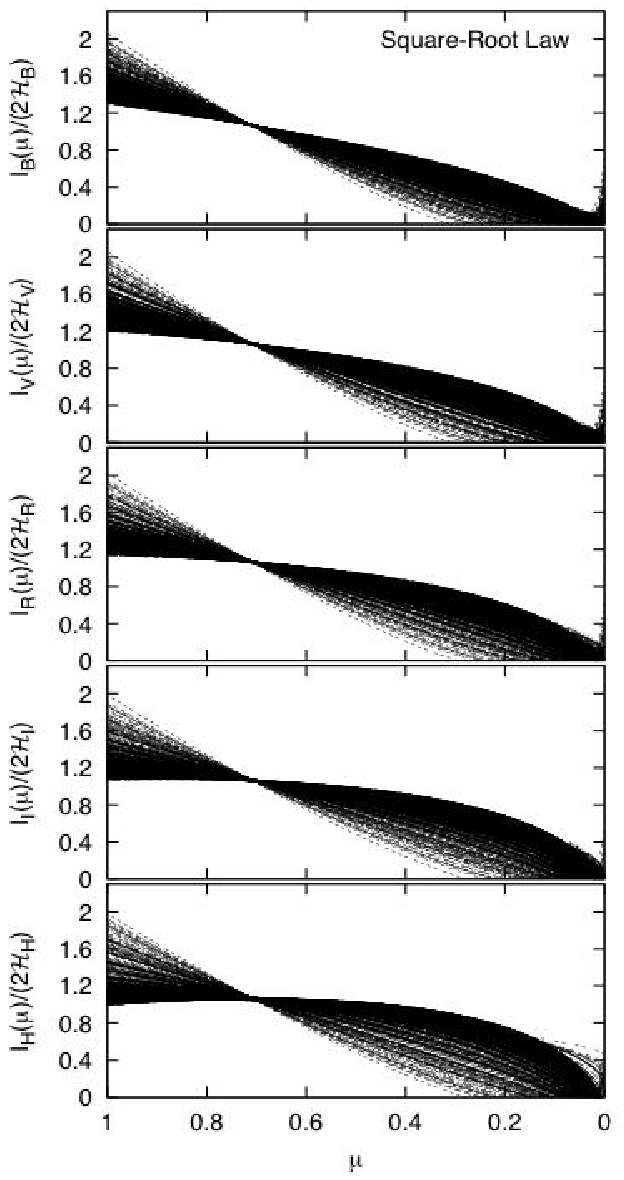}}

   \caption{Left panel: Limb darkening, characterized by 
            the normalized surface intensity, $I(\mu)/2\mathcal{H}$, 
            computed for the cube of 2102 model atmospheres for 
            $v_\mathrm{turb} = 2$ km/s for the $BVRI$ and $H$-bands.  
            Right panel: Fit to these same intensities 
            using the limb-darkening law 
            \mbox{$I(\mu)/2\mathcal{H} = 1 - A (1 - \frac{3}{2} \mu)
               - B (1 - \frac{5}{4} \sqrt{\mu})$}
            for  the same wavebands.
            \label{fig:ld_cubev2_mu}
           }
\end{figure*}
The number of models in each cube is large.  For example, there were 
2101 models in the cube with $v_\mathrm{turb} = 2$ km/s and the cubes 
for the other values of $v_\mathrm{turb}$ have similar numbers of 
models.  For each model in each cube we computed the intensity at 1000 
equally spaced $\mu$ points at all wavelengths from the far ultraviolet 
to the far infrared; the results are shown in the left column of 
Fig.~\ref{fig:ld_cubev2_mu} for the $B,V,R,I$ and $H$-bands.  These intensity curves have similar behavior as the \textsc{Phoenix} model atmosphere intensity curves that \cite{Orosz2000} studied.
We also fit each of the model limb-darkening curves with the 
linear-plus-square-root limb darkening law given by 
Eq.~\ref{eq:fields}, and this is shown in the right column of 
Fig.~\ref{fig:ld_cubev2_mu}.  Note that the $A$ and $B$ coefficients of 
this parametrization have been shown in Eq.~\ref{eq:lsa2} and 
Eq.~\ref{eq:lsb2} to be functions of $J$ and $\mathcal{P}$, not the 
structure of the intensity profile.  The results for the 
$v_\mathrm{turb} = 0$ and $4$ km/s cubes of models are very similar.  This is not surprising; in the \textsc{Atlas} codes, the turbulent pressure is $P_{\rm{turb}} = \rho v_{\rm{turb}}^2$, indicating that the turbulent velocity has a similar effect on the temperature structure of a stellar atmosphere as the gravity, that is a change of turbulent velocity is equivalent to a change of gravity, similar to the discussion from \cite{Gustafsson2008} for the MARCS code. In terms of the best-fit relations, a change of turbulent velocity causes a small change for the coefficients in the same manner as a change in gravity for the same effective temperature and mass.  Therefore, we can just explore the grid of models atmospheres for only one value of $v_{\rm{turb}}$.
The cubes of models used in this survey cover a much wider range of 
atmospheric parameters than was used in the only previous 
investigation using spherical model stellar atmospheres 
\citep{2003A&A...412..241C}, which also truncated the profiles at the 
limb to achieve better fits.

The model intensities in the left column of Fig.~\ref{fig:ld_cubev2_mu} 
are always positive, but the intensities in some of the parametrized 
fits in the right column become slightly negative toward the limb.  
This does not happen for plane-parallel model atmospheres because the 
slope of the intensity profile ($\partial I/\partial \mu$ or 
$\partial I/\partial r$) has a constant sign.  For example, using 
$\mu$ as the independent variable the slope of the intensity profile is 
always positive, varying from zero at the center of the disk to 
$\infty$ as $\mu \rightarrow 0$. For spherical atmospheres, the 
center-to-limb variation changes sign because of a slight inflection, 
which is apparent in the curves of the left column of 
Fig.~\ref{fig:ld_cubev2_mu}.  The more complex intensity profile and 
the use of an equal-weighting $\chi^2$-fit (i.e. the same number of 
$\mu$-points near the center as near the limb) leads to some slightly 
negative intensities in some best-fit relations.


Figure~\ref{fig:ld_cubev2_mu} shows that the actual limb darkening 
computed from the spherical models is more complex than the 
parametrized representation, and it is obvious that the analysis of 
specific observations that contain intensity information should be 
cautious about using the parametrized fits.  Although fits to 
plane-parallel models appear to be very good, the models are less 
realistic physically than the spherical models.  Therefore, using what 
seems like a well-fitting law may introduce hidden errors that could 
compromise the conclusions of the analysis. The quantities derived 
from observations of binary eclipses and planetary transits may be more 
uncertain than had been thought 
\citep[see, for example,][]{2007ApJ...655..564K, 2007A&A...467.1215S, 
2008MNRAS.386.1644S}.

\begin{figure}[t]
\begin{center}
 \resizebox{\hsize}{!}{\includegraphics{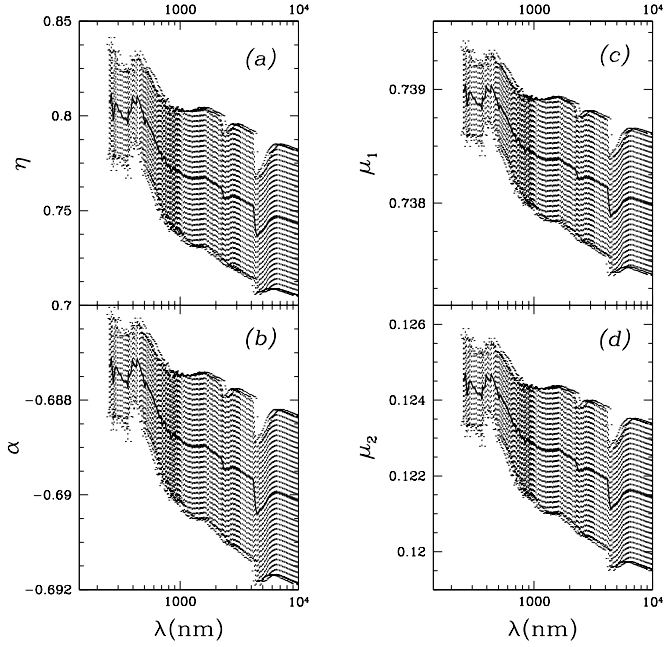}}
\end{center}
\caption{Wavelength dependence of the mean values of
         (a) the ratio of the pseudo-moment to the mean intensity,
             $\eta$,
         (b) the value of $\alpha$ that defines the fixed points,
         (c) the first and 
         (d) second fixed point for the grid of spherically symmetric 
         \textsc{Atlas} model atmospheres. The error bars represent 
         1-$\sigma$ deviation from the mean.
         \label{fig:wl_sp}
        }
\end{figure}

Using the same grid of spherical \textsc{Atlas} models used in 
Fig.~\ref{fig:ld_cubev2_mu}, we computed the mean values and 1-$\sigma$ 
standard deviations of $\eta$, $\alpha$ and the fixed points as a 
function of wavelength.  The results are shown in Fig.~\ref{fig:wl_sp}, 
which can be compared with the plane-parallel results shown in 
Fig.~\ref{fig:wl_pp}.  There are two key differences between the 
plane-parallel and spherical models. The first is that the mean values 
of $\eta$ are larger for the spherical models, which leads to 
the differences between the mean values of $\alpha$ and the two fixed 
points.  The second difference is that the standard deviation for 
$\eta$ as a function of wavelength is much larger for the spherical 
models than for the plane-parallel models.  This suggests that $\eta$ 
varies much more as a function of effective temperature, gravity and 
mass. Clearly, the fixed point is much more constrained in 
plane-parallel model stellar atmospheres. 

\begin{figure}
   \resizebox{\hsize}{!}{\includegraphics{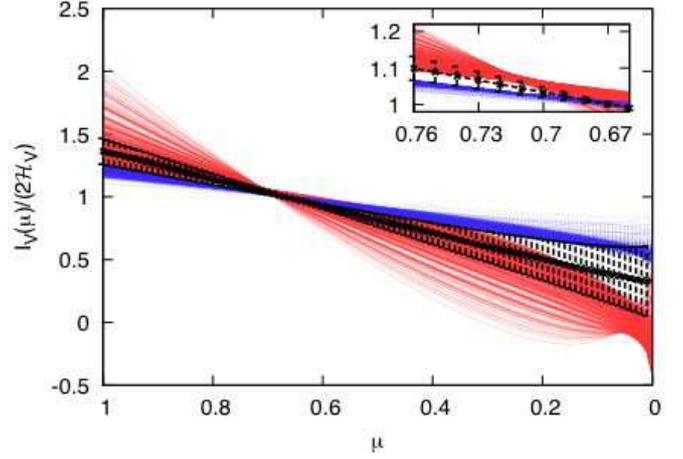}}
   \caption{Comparison of the $V$-band limb-darkening relation 
            determined by \citet{2003ApJ...596.1305F} from 
            observations (black dashed curves with errorbars) with the limb-darkening 
            relations computed from \textsc{SAtlas} model stellar 
            atmospheres (red solid curves) and relations computed from \textsc{Atlas} model stellar atmospheres (blue dotted curves), represented using the same 
            linear-plus-square-root law.  The insert shows a magnified 
            view of the region of the fixed $\mu$-point.
            \label{fig:vld}
           }
\end{figure}

A specific example of an application of limb darkening to an 
observation is the analysis by \citet{2003ApJ...596.1305F} of 
microlensing observations.  In Fig.~\ref{fig:vld}, we compare the 
$V$-band limb-darkening relations from our cube of spherical models 
with the $V$-band limb-darkening relation determined by 
\citet{2003ApJ...596.1305F} from observations.  It is clear that the 
\citet{2003ApJ...596.1305F} $V$-band limb-darkening law agrees well 
with the limb-darkening from the spherical models, but, not as well with the plane-parallel models.  Furthermore, 
although the curves for the model stellar atmospheres narrow to a 
waste rather than to a point, the location of minimum spread coincides 
with the location of the observed $\mu$-position in the $V$-band 
limb-darkening relation within the uncertainty of the observed 
coefficients.  Even though the uncertainty of limb-darkening relations 
from microlensing observations is large, these results suggest that the 
\citet{2003ApJ...596.1305F} limb-darkening observations are probing the 
extended the atmosphere of the lensed red giant star.
\begin{figure}
  \resizebox{\hsize}{!}{\includegraphics{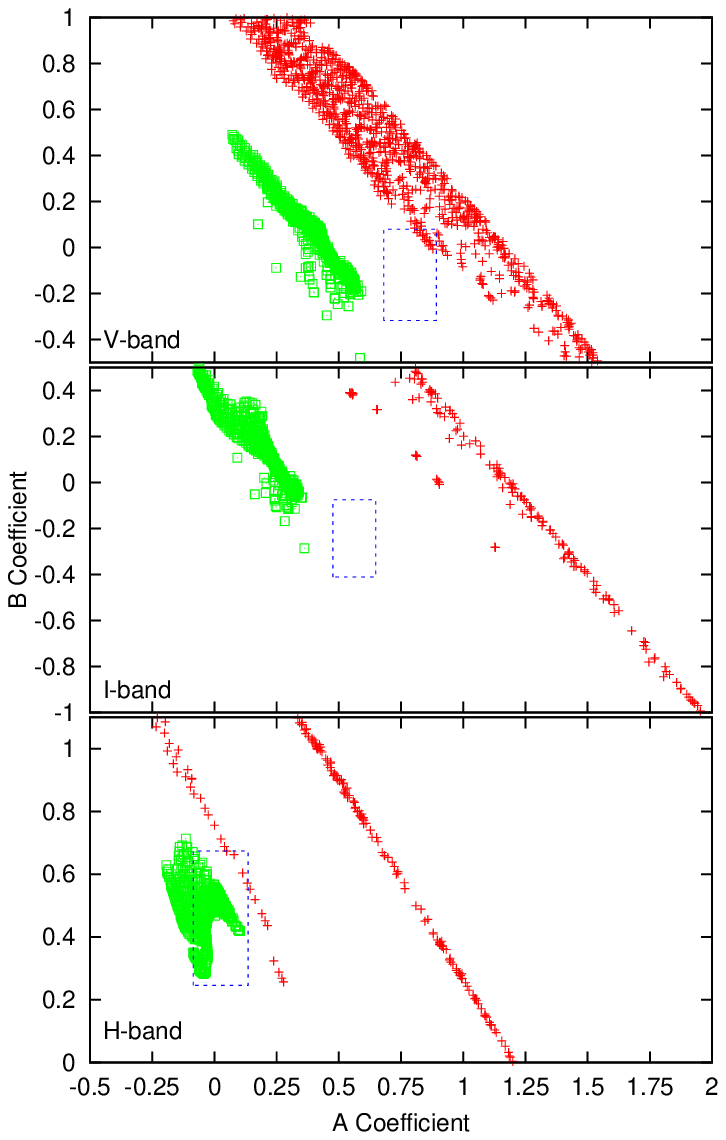}}
  \caption{Red plus symbols represent the limb-darkening coefficients $A$ 
           and $B$ computed from \textsc{SAtlas} model intensity 
           profiles for the $V$, $I$, and $H$-bands, and green open squares 
            represent the coefficients computed with 
           plane-parallel \textsc{Atlas} models.
           The box in each plot shows the range of the 
           coefficients derived from microlensing observations by 
           \citet{2003ApJ...596.1305F}. 
           \label{fig:abbox}
          }
\end{figure}
We conclude that the parametrized laws, although more simplified than 
the computed limb-darkening curves, are useful for understanding how 
limb darkening depends on the fundamental properties of the stellar 
atmosphere.

To explore further the information content of the limb darkening, we 
recall that Eq.~\ref{eq:a_fn_b} showed that the coefficients $A$ and 
$B$ of the parametrization are linearly correlated.  In 
Fig.~\ref{fig:abbox} we plot the limb-darkening coefficients $A$ and 
$B$ for the $V$, $I$, and $H$-bands computed with both the 
plane-parallel \textsc{Atlas} and the spherical \textsc{SAtlas} models.
For each wavelength the error box shows the range of $A$ and $B$ found 
by \citet{2003ApJ...596.1305F} from their observations.  For the 
$V$-band, the spherical limb-darkening coefficients overlap with the 
observed fit, but the plane-parallel coefficients do not, while for 
$H$-band both plane-parallel and spherical models agree with 
observations.  The microlensing observations at the longer wavelength 
might not be sensitive enough to probe the low intensity limb of the 
star, making the star appear consistent with limb-darkening of the 
plane-parallel model atmospheres.   In the $I$-band neither set of 
models agree with the observations, but the $I$-band data provide a 
weak constraint.  As \citet{2003ApJ...596.1305F} noted, the $I$-band 
time-series microlensing observations were a composite from multiple 
sites, and removing data from any one site changed the results 
significantly.
\begin{figure*}
  \resizebox{\hsize}{!}{\includegraphics{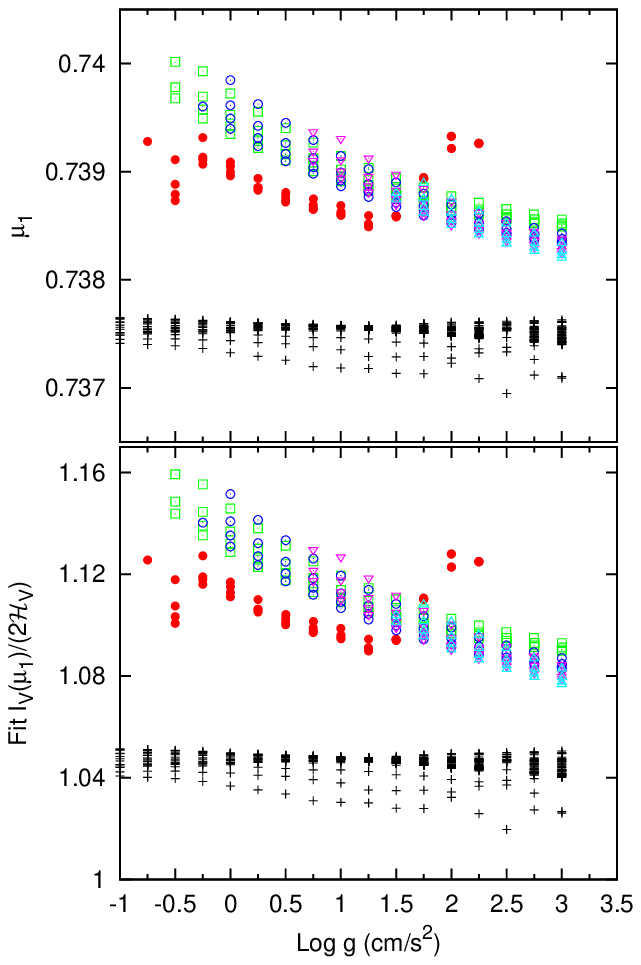}
                        \includegraphics{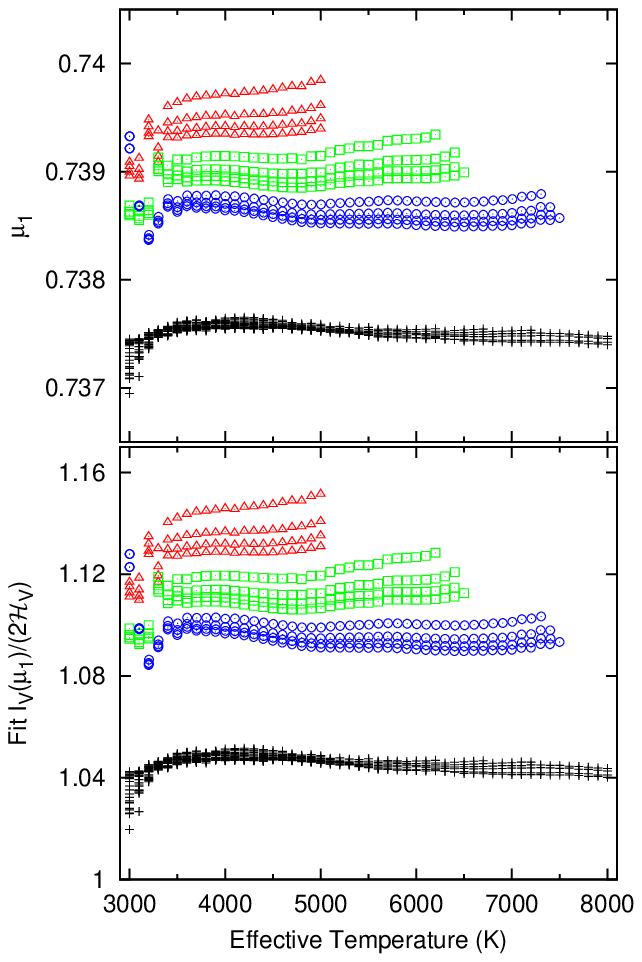}}
  \caption{Left: As a function of surface gravity, the top panel 
           shows the value of the primary fixed point, $\mu_1$, for 
           the linear plus square root parametrization, for spherical 
           atmospheres of varying effective temperature. 
           The bottom panel shows the dependence of the normalized 
           V-band intensity of the fixed point.  
           Red fiilled circles are $T_\mathrm{eff} = 3000$ K, green open squares 
           4000 K, blue open circles 5000 K, magenta downward pointing 
           triangles 6000 K and pale blue upward pointing triangles are 
           7000 K.  The black crosses show the behavior of the fixed point from the grid of plane-parallel model atmospheres for comparison
           Right: As a function of effective temperature, the top 
           panel shows the fixed point, $\mu_1$, and the bottom panel 
           shows the dependence of the V-band intensity of the fixed 
           point. 
           Red triangles represents models with $\log~g = 0$, green squares 
           $\log~g = 1$, and blue circles are $\log~g= 2$. Again, black crosses represent the fixed point from plane-parallel model atmospheres.
           \label{fig:gravity_dep}
          }
\end{figure*}

In the $V$-band, the limb-darkening coefficients from 15 model 
atmospheres fall within the observational box.  These models have 
$\log~g = 2.25 - 3$ and $T_\mathrm{eff} = 3400 - 3600$ K.  For the 
$H$-band data there are four models within the observational box; 
these have same range of gravities but are slightly cooler, 
$T_\mathrm{eff} = 3000 - 3100$ K.   It is interesting that the models 
that agree with the observations are those with gravities that are 
consistent with the results of \cite{2003ApJ...596.1305F} and 
\cite{2002ApJ...572..521A}.  This suggests that observations using the 
limb-darkening parametrization can probe the spherical extension of 
stellar atmosphere via the fixed point, $\mu_1$, and potentially probe 
the fundamental parameters of stars via the limb-darkening coefficients.

Using a cube of models with a given value of $v_\mathrm{turb}$, we 
examine the dependence of the primary fixed point, $\mu_1$, on the 
effective temperature and surface gravity.  Note again that there are 
three basic parameters characterizing the spherical atmospheres: 
$L_\star$, $M_\star$ and $R_\star$.  This means that values of 
$T_\mathrm{eff}$ and $\log g$ are degenerate, but they are easier to 
show on a two-dimensional surface.  In the top left panel of 
Fig.~\ref{fig:gravity_dep} we plot the value of the primary fixed 
point, $\mu_1$, as a function of $\log g$.
At each value of $\log g$ there are values of $T_\mathrm{eff}$ 
ranging from 3000 to 7000 K, although there can be multiple values 
because of the parameter degeneracy.  In the bottom left panel of 
Fig.~\ref{fig:gravity_dep} we plot 
$I_\mathrm{V}(\mu_1)/2\mathcal{H}_\mathrm{V}$, the normalized 
intensity of the fixed point in the $V$-band.  For each surface 
gravity the values of $\mu_1$ and the normalized intensity at the 
fixed point both show a steady progression as $T_\mathrm{eff}$ 
changes, except for  $T_\mathrm{eff} = 3000$ K.  We suspect that the 
behavior for $T_\mathrm{eff} = 3000$ K is due to a change in the 
dominant opacity source for our coolest models, possibly water vapor.  Models with 
$T_\mathrm{eff} > 3000$ K have H$^-$ as the dominant continuous 
opacity, but in the coolest models there are fewer free electrons 
available to form H$^-$, and the formation of H$_2$ reduces the pool of 
hydrogen atoms.  We also plot the fixed point and intensity at the fixed point from plane-parallel model atmospheres for comparison.  It is clear that the fixed point from fits to plane-parallel models varies much less than fits to spherical models.

On the right side of Fig.~\ref{fig:gravity_dep} we reverse the 
parameters and plot the dependence of the primary fixed point and the 
normalized intensity of the fixed point as a function of 
$T_\mathrm{eff}$.  At each value of the effective temperature, the 
values of $\log g$ are 0, 1 and 2 in cgs units.  In the top right panel 
of Fig.~\ref{fig:gravity_dep} we see that there is essentially no 
variation of the value of $\mu_1$ with $T_\mathrm{eff}$ for all three 
surface gravities until the lowest effective temperature is reached.  
There is an obvious displacement of the value of $\mu_1$ for each 
gravity and also a spread in $\mu_1$ because of parameter degeneracy.  
However, for $T_\mathrm{eff} \leq 3500$ K the value of $\mu_1$ drops 
for all surface gravities.  The bottom right panel shows that the 
normalized $V$-band intensity of the fixed point shows a similar 
behavior.  For $T_\mathrm{eff} > 3500$ K there is little variation 
with $T_\mathrm{eff}$, but there is an offset and a spread that 
depends on the surface gravity.  Cooler than 3500 K the value of the 
normalized intensity drops noticeably.

It is clear that the fixed point $\mu_1$ and the intensity at the fixed point $I(\mu_1)/2\mathcal{H}$ are functions of gravity and effective temperature for spherically-symmetric models and are roughly constant for plane-parallel model atmospheres.  This indicates that the best-fit coefficients of the limb-darkening law vary mostly because of the geometry of the model atmosphere.  A spherical model atmosphere predicts a smaller intensity near the limb of the stellar disk relative to a plane-parallel model with the same effective temperature and gravity.  To predict the same emergent flux, the intensity must be larger at the center of the disk, hence the temperature at the base of the atmosphere must also be larger for the spherical model.  Therefore, the temperature structure of the atmosphere also varies.  However, this is a secondary effect and the geometry of the atmosphere is most important in determining the value of the $\mu_1$ and $I(\mu_1)/2\mathcal{H}$.  The geometry of spherical models leads to smaller values of the pseudo-moment and the mean intensity because the intensity is more centrally concentrated.  This suggests that  $\mu_1$ and $I(\mu_1)/2\mathcal{H}$ depend on the atmospheric extension. We will explore how the fixed point and intensity relate to the extension and fundamental stellar properties in greater detail in a future article.

\section{Conclusions}

We have explored limb darkening using large cubes of spherical stellar 
atmospheres spanning the parameters $L_\star$, $M_\star$ and $R_\star$ 
covering the cool, luminous quadrant of the Hertzsprung-Russell 
diagrams (Fig.~\ref{fig:ld_hr}).  These models have also used three 
different values of the microturbulent velocity.  For each model, the 
center-to-limb variation of the surface intensity has been calculated 
at 1000 equally spaced $\mu$ values spanning the range from 1 to 0 for 
every wavelength used to compute the model structure.

Parametrizing the center-to-limb variation with a flux-conserving 
linear-plus-square-root limb-darkening law, we confirm the findings of 
\citet{2000PhDT........11H} and \citet{2003ApJ...596.1305F} that there 
is a fixed $\mu_1$ point through which all the intensity curves pass.
However, when we plot the surface intensities directly, without using 
a fitting law, there is no fixed point, although the distribution of 
curves does narrow to a waist close to the same value of $\mu_1$ 
(Fig.~\ref{fig:ld_cubev2_mu}).

The apparent fixed point is a result of the least-squares fitting 
procedure where the two parameters of the law are dependent on two 
properties of the stellar atmosphere, the mean intensity, $J$, and the 
pseudo-moment, $\int I \sqrt{\mu} \mathrm{d} \mu$.  For the temperature 
range $4000 - 8000$ K, the mean intensity is correlated with the 
pseudo-moment, which means that the two coefficients are also 
correlated, leading to the existence of the fixed point.

The lack of a well-defined fixed point in the surface intensity 
distribution for spherical model atmospheres suggests that the three 
fundamental parameters of the atmospheres affect the limb darkening in 
a way that is not encountered in the two-parameter plane parallel 
atmospheres.

\acknowledgements

This work has been supported by a research grant from the Natural 
Sciences and Engineering Research Council of Canada.  HRN has received 
financial support from the Walter John Helm OGSST, the Walter C. 
Sumner Memorial Fellowship, and the Alexander von Humboldt Foundation.

\bibliographystyle{aa} 

\bibliography{ld_p1_v2} 

\end{document}